# Four principles for improved statistical ecology

Gordana Popovic[1,6], Tanya J. Mason[2], Tiago A. Marques[3,4], Joanne Potts[5], Szymon M. Drobniak[6,7], Rocío Joo[8], Res Altwegg[9], Carolyn C. I. Burns[10], Michael A. McCarthy[11], Alison Johnston[12], Shinichi Nakagawa[6], Louise McMillan[13], Kadambari Devarajan[14,15], Patrick l. Taggart[16], Alison C. Wunderlich[17], Magdalena M. Mair[18,19], Juan Andrés Martínez-Lanfranco[20], Malgorzata Lagisz[6], Patrice P. Pottier[6]

1. Stats Central, Mark Wainwright Analytical Centre, UNSW Sydney, Australia
2. Centre for Ecosystem Science, School of Biological, Earth and Environmental Sciences, UNSW Sydney, Australia
3. Centre for Research into Ecological and Environmental Modelling, The Observatory, University of St Andrews, St Andrews, Scotland
4. Centro de Estatística e Aplicações, Departamento de Biologia Animal, Faculdade de Ciências da Universidade de Lisboa, Portugal
5. The Analytical Edge Statistical Consulting, PO Box 47, Blackmans Bay, Tasmania, Australia
6. Evolution and Ecology Research Centre, School of Biological Earth and Environmental Science, University of New South Wales, Sydney, Australia
7. Institute of Environmental Sciences, Jagiellonian University, Krakow, Poland
8. Global Fishing Watch, Washington, DC 20036, USA
9. Centre for Statistics in Ecology, Environment and Conservation, Department of Statistical Sciences, University of Cape Town, 7701 Rondebosch, South Africa
10. Sydney, Australia
11. School of Ecosystem and Forest Sciences, The University of Melbourne, Parkville, Victoria, Australia
12. Centre for Research into Ecological and Environmental Modelling, Mathematics and Statistics, University of St Andrews, St Andrews, UK
13. School of Mathematics and Statistics, Victoria University of Wellington, Wellington, New Zealand
14. Organismic and Evolutionary Biology Graduate Program, University of Massachusetts at Amherst, Amherst, MA, USA
15. Department of Natural Resources Science, University of Rhode Island, Kingston, RI, USA
16. Vertebrate Pest Research Unit, Department of Primary Industries NSW, Queanbeyan, New South Wales, Australia
17. Institute of Biosciences, São Paulo State University, Coastal Campus, São Vicente, São Paulo, Brazil
18. Statistical Ecotoxicology, University of Bayreuth, Bayreuth, Germany
19. Theoretical Ecology, University of Regensburg, Regensburg, Germany
20. Department of Biological Sciences, University of Alberta. Edmonton. Alberta, Canada

# Abstract

Increasing attention has been drawn to the misuse of statistical methods over recent years, with particular concern about the prevalence of practices such as poor experimental design, cherry-picking and inadequate reporting. These failures are largely unintentional and no more common in ecology than in other scientific disciplines, with many of them easily remedied given the right guidance.

Originating from a discussion at the 2020 International Statistical Ecology Conference, we show how ecologists can build their research following four guiding principles for impactful statistical research practices: 1. Define a focused research question, then plan sampling and analysis to answer it; 2. Develop a model that accounts for the distribution and dependence of your data; 3. Emphasise effect sizes to replace statistical significance with ecological relevance; 4. Report your methods and findings in sufficient detail so that your research is valid and reproducible.

Listed in approximate order of importance, these principles provide a framework for experimental design and reporting that guards against unsound practices. Starting with a well-defined research question allows researchers to create an efficient study to answer it, and guards against poor research practices that lead to false positives and poor replicability. Correct and appropriate statistical models give sound conclusions, good reporting practices and a focus on ecological relevance make results impactful and replicable.

Illustrated with an example from a recent study into the impact of disturbance on upland swamps, this paper explains the rationale for the selection and use of effective statistical practices and provides practical guidance for ecologists seeking to improve their use of statistical methods.

# Introduction

When reporting research findings, ecologists, like other scientists, want their results to reflect what truly happens in the system being studied, and communicate both ecological relevance and the level of evidence. For their results to hold up, researchers need to follow good research practices. Not following good practices has led to low reproducibility of findings in many fields (Open Science Collaboration, 2015; Begley & Ellis, 2012; Camerer et al., 2018). Poor research practices are also common in ecology (Anderson et al., 2000; Fidler et al., 2006; Fraser et al., 2018), and their use can distort findings, waste resources, inadequately report what is happening in ecological communities, and ultimately have the potential to misrepresent research to other researchers, policy makers and the public.

Poor research practices stem mostly from misunderstandings or misinterpretations of statistical methods and principles of study design. Some of the most common and consequential of these practices include:

- Hypothesising after results are known (HARKing; Kerr, 1998), a practice that 51% of ecologists and evolutionary biologists report engaging in (Fraser et al., 2018).
- Not reporting non-significant results; a form of cherry-picking, which 64% of ecologists admitted to doing at least once (Fraser et al., 2018).
- Hypothesis testing based on a null hypothesis that is known a priori to be false. Anderson et al., 2000 found the vast majority (95%) of *Ecology* articles they evaluated contained null hypotheses that were likely known to be false a priori.
- Misinterpreting non-significant results as evidence of "no effect" or "no relationship"; which happens approximately 63% of the time non-significant results are reported in ecology (Fidler et al., 2006).
- Providing insufficient detail on methods and analysis. Almost 80% of ecology papers do not provide enough detail to be computationally reproducible (with most, 73%, failing to include accompanying analysis code; Culina et al., 2020).

When presented with this report card, ecologists may justify their research practices as being relatively harmless or unavoidable. However, misinterpretations of the role of hypothesis tests makes any such tests and their results meaningless. In addition, cherry-picking and HARKing are known to (vastly) inflate the rate of false positives ("significant" results when there is no actual effect in the population, see for example Forstmeier et al., 2017), leading to publication of many incorrect results. Unfortunately, we do not know the true rate of false positives in ecology, as replication studies are very uncommon, with only 0.023% of all studies published to date representing a true replication (Kelly, 2019), even while ecologists and evolutionary biologists universally claim that replication is very or somewhat important (97% of respondents; Fraser et al., 2018). However, we do know that an uncomfortably large proportion (70%) of studies in ecology support their original hypotheses (Fanelli, 2010), suggesting many of these are false positives. The other side of the coin is research waste. Purgar et al., 2021 found that between 82% and 89% of research in ecology appears to be avoidably wasted due to a combination of low-quality studies, publication bias and poor study design, analysis, and reporting.

In framing statistical best practice with reference to four principles, we hope to guide ecological researchers, particularly those just starting out, to present the best possible evidence for their conclusions by avoiding these pitfalls. The four principles we have identified are:

1. First, define a focused research question, then plan sampling and analysis to answer it.
2. Develop a model that accounts for the distribution and dependence of your data.

3. Emphasise effect sizes to replace statistical significance with ecological relevance.
4. Report your methods and findings in sufficient detail so that your research is valid and reproducible.

These principles are listed in approximate order of importance and impact, and later principles require the foundation provided by earlier principles to be effective. For example, reporting effect sizes will not improve a study if the model does not account for dependence. Defining focused research questions before any data are collected can eliminate HARKing, especially when paired with registration, which has the additional benefit of eliminating cherry-picking. Developing a plan for sampling helps to better answer research questions, leading to more meaningful and impactful results, and increases the likelihood true effects are found (statistical power) with better sampling design. Correctly modelling data also increases power and controls the rate of false positives. Emphasising effect sizes puts the focus on ecological relevance, which is the most meaningful result of ecological research. Comprehensive reporting of methods and findings allows others to successfully replicate your study, lending more weight to your findings and moving the field forward.

These four principles arose out of conversations at the 2020 International Statistical Ecology Conference. The discussion group which conceptualised this paper included a range of ecological statisticians currently working in the field. We noticed that while excellent literature describing good practices in statistics exists, including in ecology, these tend to focus on protocols for conducting specific analyses (e.g.; Zuur et al., 2010; Steel et al., 2013; Zuur & Ieno, 2016) or addressing specific problems (Nakagawa & Parker, 2015), rather than a small and digestible number of principles to follow for all analyses. By defining these principles, we hope to empower ecologists to pursue more robust and meaningful research and encourage collaborations in ecological statistics by helping to develop a common research *methodology.*

Throughout, we will demonstrate the principles with an ecological example examining how disturbances affect upland swamps (Mason et al., 2022). Underground mining is known to disrupt surface and groundwater flows which may affect nearby swamp communities. The researchers wanted to examine how differing water availability affected swamp plant communities, both alone and in combination with a fire disturbance. For this study, mesocosms were collected from multiple swamps, then randomised to water and fire treatments in a glasshouse. Mesocosms, for the purpose of this study, are a column of soil and plants, collected by hammering PVC pipe (diameter of 150 mm and a depth of 250 mm) to ground level and extracted with trenching shovels. They were then placed in tubs in a glasshouse, and tub water levels were manipulated to simulate different levels of groundwater availability. A fire event was simulated by sequentially applying biomass removal (clipping), heat and smoke to half of the mesocosms in each water treatment after 20 months (see Figure 1).

Within each section we will mention useful R packages (R Core Team, 2020) for each principle. We chose R since it is open and free software and the most used software in ecology (Lai et al., 2019).

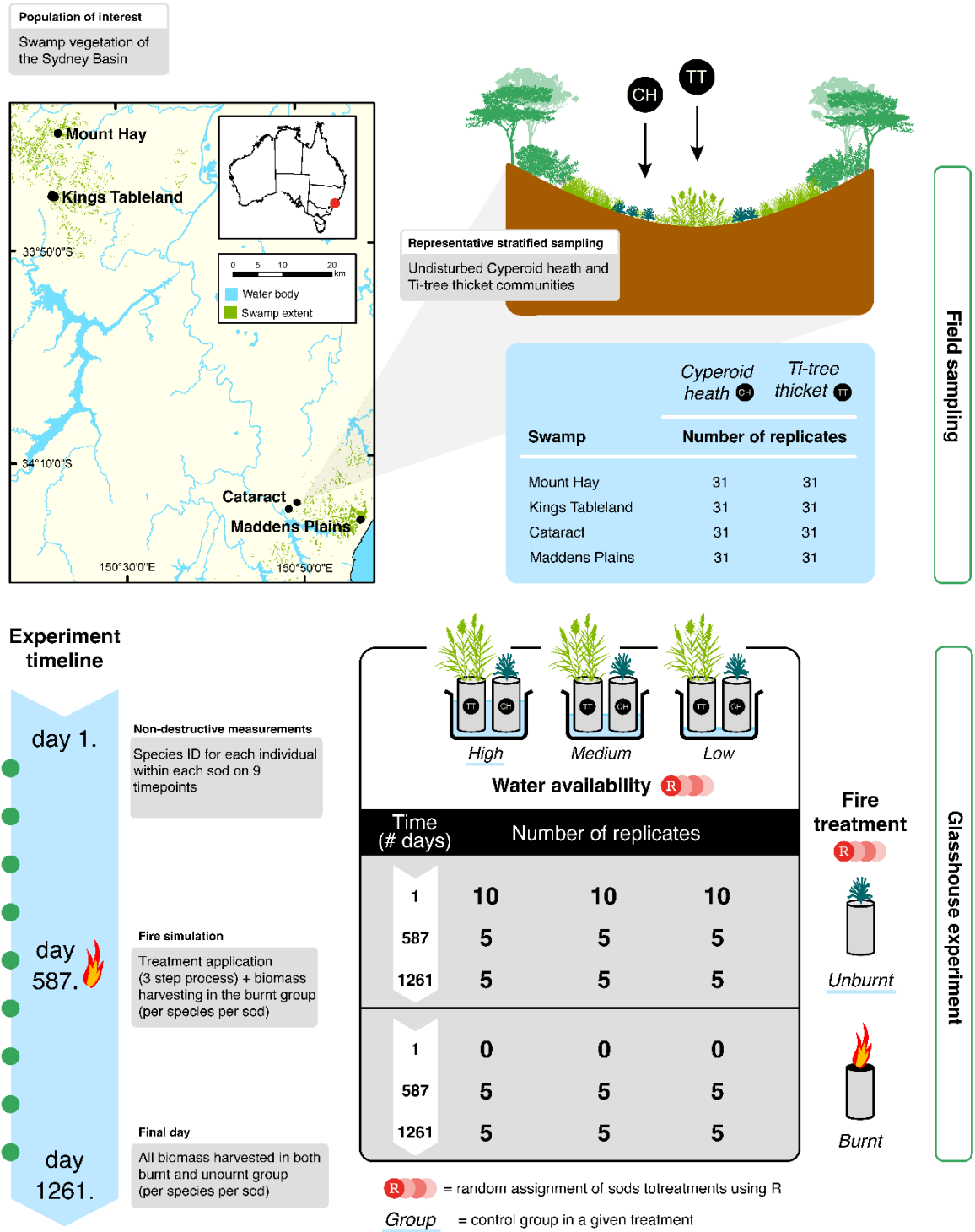

*Figure 1 – Experimental design for swamp study*

# Principle 1. First, define a focused research question, then plan sampling and analysis to answer it

## 1A – Define your research question

Developing a good research question is the most important part of the research process. A poorly conceived question can not only limit the usefulness of your findings but can also lead to methodological problems throughout the study. A good research question is Feasible, Interesting, Novel, Ethical, and Relevant (the FINER criteria; Hulley, 2007) and each research question should lead to a specific statistical analysis plan, which you would ideally register (see section 1C). In the swamp example, the researchers considered a glasshouse experiment to be feasible and ethical, literature review confirmed it was novel and it was both interesting and relevant as planning approval for longwall mining under swamps is a live and highly contested issue in wetland management.

The PICO framework (Haynes, 2006) is often used to frame research questions in health and medical studies, but can readily be adapted to ecological studies. For each research question, it is important to define the:
- **Population** – population of interest or the target population, usually defined by species, space, and time.
- **Intervention** – the treatment that will be applied to subjects in a randomised experiment, and the explanatory variable of interest in an observational study.
- **Comparison** – identifies what you plan to use as a reference or control group to compare with your treatment group.
- **Outcome** – represents what outcome(s) you plan to measure to examine the effectiveness of your intervention or effect of the explanatory variable of interest, often called the response. In ecology this may be abundance, richness, location, velocity, diversity, among others.

It is often helpful to try to predict what will happen in your study, as this can help clarify your research question. The swamp researchers thought that mesocosms with less available water would have lower biomass and richness than mesocosms with more water availability, and that the effect of lower water would be compounded by fire. The population of interest in this example upland swamp plant communities of the Sydney Basin, Australia. The intervention groups were combinations of water level and fire: with low, medium, and high levels of water availability; and burnt and unburnt fire levels. The control group was represented by high water mesocosms. Outcomes included biomass, richness, and presence/absence of each species (species composition).

Clearly defining a research question *before* collecting data (or exploring previously collected data) is the most important step in ensuring robust experimental design. While the benefits of clearly defining a research question are clear, it may be less obvious why it is important not to refine or revise the hypothesis after conducting your study. Hypothesising after the data relating to your original research question have been gathered has a distorting effect on your results, deceptively suggesting that the evidence for your *post hoc* hypothesis is overly strong (Forstmeier et al., 2017). This is because any changes you make to your hypothesis after collecting data are likely to reflect patterns in the sample which do not necessarily reflect patterns in your population, increasing the likelihood your results will indicate a false positive.

This prohibition against *post hoc* hypothesising does not however prevent researchers from doing exploratory analyses to inform future research questions and hypotheses, indeed this can be one of the best ways to define productive avenues for further research. As ecologists, we often summarize very complex data to model it simply. For example, you might combine multi-species data into a richness metric to answer your primary question, as in the swamp study. This complexity however is often of great interest for understanding the ecology of your study species and communities, and spending time exploring your raw data often leads to new discoveries and avenues of research. Data exploration can take the form of plotting, tabulating, model fitting and hypothesis testing (including *p*-values). Any time you explore data without a prespecified research question in mind, you should think of it (and report it) as exploratory or hypothesis generating (Forstmeier et al., 2017).

## 1B – Match data collection to research aims

Once research questions are outlined, the next step is to decide which data are needed to answer them. Collecting data may include observation, sampling, and experimental manipulation. Manipulative experiments are the only way to show causation and should always be performed when feasible. However, manipulative experiments do have disadvantages in ecology; they are limited in their scope (time constraints and number and type of treatment), and can be unethical (e.g. creating an environmental disturbance with large negative effects). They may also not be a realistic setting, for example the swamp study was limited as recruitment of new species to the mesocosms was not possible in the glasshouse. It is best to complement manipulative experiments with natural experiments and longitudinal monitoring (Driscoll et al., 2010).

**Sampling**

Representative samples will reflect the patterns and characteristics of the target population. In our swamp example, the target population is composed of species from the wettest sub-communities (ti-tree thicket and cyperoid heath) of undisturbed upland swamp communities of the Sydney Basin.

When collecting new data in the field, a representative sample is the best way to ensure unbiased inference. The complex spatial and temporal nature of ecology can make representative sampling designs challenging, and ecologists have in the past settled for less representative or even haphazard data collection (Smith et al., 2017). We advocate for ecologists to use representative samples to produce the most reliable and unbiased results. Helpful tools are the `sample()` function in base R, and the `spsample()` function in the `sp` package (Pebesma & Bivand, 2005).

When representative sampling is not achieved due to unexpected challenges in the field, or biased pre-existing datasets, it may be possible to account for the non-representative sampling in modelling by using weights and offsets (to correct for known biases, e.g. sampling intensity), covariates (to correct for imbalances associated with measured variables, e.g. site accessibility), and correlation structures (See Principle 2A). Alternatively, you can frame your inferences to the population for which the sample you have is representative (Williams & Brown, 2019). When working with pre-existing datasets rather than collecting new field data, it is important to consider any inherent biases. When using data collected by others, it is good practice to contact and collaborate with the researchers that collected the data, as they have the best understanding of their data. A representative sampling design can be approximated by carefully selecting the data to use in analyses (Johnston et al., 2021).

**Experiments, correlation, and causation**

Researchers are most often interested in causal relationships, to answer questions about if and how a response is promoted, caused, or induced by a given set of covariates. Finding such cause-and-effect relationships might be the first step to unveil possible ecological mechanisms. The only way to definitively demonstrate causation is via experiments, where researchers manipulate the environment and observe the consequences. Experiments must have:
1. Controls (to know what would have happened without the manipulation)
2. Replication (within treatments; to apportion observed differences to the manipulation instead of random variation)
3. Randomisation (of sampling units to treatment; to avoid bias and confounding with unmeasured / uncontrolled variables)

The swamp example demonstrates how this can be done even in complex environments. Mesocosms were collected from several swamps, then randomised to treatments, allowing the researchers to demonstrate the cause-effect relationships between water and fire treatments and plant communities. There were several treatment combinations, with the high water unburnt mesocosms acting as the control. Replication was achieved by sampling and allocating multiple mesocosms to each treatment. Sampling units (mesocosms) needed to be allocated to treatments (e.g. low, medium, and high water levels) randomly. This randomisation can easily be achieved using widely available software like the `randomizr` package (Coppock, 2019), as was done for the swamp example, or simply by drawing numbers out of a hat.

An additional component of many well-conducted experiments is the practice of blinding, where researchers collecting measurements or analysing data do not know the treatment allocation. Blinding can remove unconscious observer bias, but is unfortunately not widely practiced in ecology, with only about 13% of eligible ecological studies actually undertaking blinding (Kardish et al., 2015).

For ecologists, particularly field ecologists, it is often not feasible to conduct controlled experiments. The next best option is to control for confounding variables in a regression or use other 'causal' analysis methods (e.g. Larsen et al., 2019). However, results from regressions and causal analyses should only be interpreted causally with qualification as they rely on very strong assumptions of the data which can generally not be checked. Any use of causal methods should be accompanied with detailed discussion of all assumptions and their plausibility.

## 1C – Plan analysis and consider registration

Before you collect or explore the available data, it is critical to develop a robust analysis plan. Planning your approach to analysis early can substantially reduce the complexity of the final analysis and improve the clarity of your results. An effective data analysis plan ensures that your study addresses the research question and is developed in tandem with your sampling method (see Principle 2 for how to best develop a model that accounts for the characteristics of the data).

An analysis plan must include a comprehensive list of models and tests to be done, specifying in each case the model type (e.g. mixed effects model with gaussian distribution), outcome/response variable (e.g. log of richness), the principal effect of interest (e.g. water level) and variables to control for (e.g. fixed effects of swamp and vegetation community, random effect of mesocosm). Often, we will not know which model will best fit our data, in which case we should include a description of how a good model will be chosen. In the swamp example, the researchers assume a Gaussian distribution for richness to start with, then check residual plots and change to a log + 1 transformation if the Gaussian distribution is inappropriate. When conducting multiple tests, appropriate methods for controlling for multiple testing should be implemented (see e.g. Pike, 2011) to control for false positives.

For the most robust experimental design, we recommend registering your study (we follow Rice & Moher, 2019 to prefer the term registration to preregistration). Registration archives a detailed description of one's study *before* data collection or sometimes before data analysis (Nosek et al., 2018; Parker et al., 2019; Rice & Moher, 2019). Such description includes research aims (Principle 1A: research questions, hypotheses, and predictions), study design (Principle 1B; data collection process) and a data analysis plan (Principle 2A: the statistical models to be fitted). Registration can be time-stamped using public registries, such as *Open Science Framework* or *As Predicted*, and it can also be embargoed if needed. A registered report is similar to a registration in that they both commit to their study hypotheses and analysis plans prior to data collection, but it is a distinctive procedure (Chambers, 2013) where the introduction and methods are peer reviewed ahead of data collection. Currently, a registered report is considered in ecological journals, such as *BMC Ecology and Evolution* and *Ecology and Evolution* and *Conservation Biology*, and in less-specialised journals, such as *PloS One*, *BMC Biology, Scientific Reports* and *Nature Communications*.

Registration can have a substantial impact on research quality. To take an example from another field, a recent study has shown that registered reports in biomedical and psychological research supported only around 40% of their original hypotheses (Allen & Mehler, 2019) relative to 80% - 95% in traditional literature. Although we are yet to have an ecological counterpart of this study, we anticipate registration and registered reports would bring a more reasonable ratio between positive and negative findings in ecology.

# Principle 2. Develop a model that accounts for the distribution and dependence of your data

## 2A – Model dependence

Independence of errors is a critical assumption of all statistical methods. Errors are the unmodelled portion of the data, after modelling dependence and impacts of covariates. The independence assumption can be met by collecting independent data (e.g. using a simple random sample), or by appropriately modelling (accounting for) any dependence in the data.

Dependence often arises from sampling design. Consider, for instance, a hierarchical (or nested) sampling design like in the swamp example, where each experimental unit (swamp) was subsampled. Dependence arises because mesocosms within each swamp are more similar to each other than they are between swamps. Such subsampling (or 'pseudo-replication' as it is often called), is a common source of dependence (Hurlbert, 1984), as is temporal dependence whereby, for example, the abundance of species in a mesocosm is measured repeatedly. Again, these repeated measures (Gurevitch & Chester, 1986) result in observations that are more similar to each other for each mesocosm than they are between mesocosms. Spatial dependence, where sites closer in space are more similar than sites further apart, is also very common. Dependence can also arise due to the complex nature of the system. For example, in multi-species surveys, dependence between species abundances may arise from common responses to unobserved environmental gradients, phylogeny, or interactions between species.

Dependence is not problematic, per se, and replicate observations can be extremely valuable if understood, but must be accounted for correctly in the analysis (Steel et al., 2013). If dependence is not correctly accounted for, it will lead to incorrect inference (often false positives). Modelling dependence is straightforward in most statistical software (see Warton, 2022 for a detailed guide). Briefly, generalised linear (mixed) models (McCulloch & Neuhaus, 2006), implemented in the `lme4` (Bates et al., 2015, p. 4) package, can model dependence due to clustering by including a random intercept for the clustering variable ( `~ (1|mesocosm)` ). For multivariate dependence there are hierarchical models implemented in the `gllvm` (Niku et al., 2020) and `Hmsc` (Tikhonov et al., 2020) packages, as well as generalised estimating equations (e.g `mvabund`; Wang et al., 2012). For the swamp example, there are several layers of dependence. Mesocosms within one swamp are dependent, repeated measurements of each mesocosm are dependent and measurements include multiple species, which have complex interdependences. The swamp researchers fitted hierarchical models using the `glmmTMB` (Brooks et al., 2017), where in addition to the effects of interest (vegetation, water and fire treatments), they included fixed effects for swamp, random effects for mesocosm and a reduced rank correlation structure between species (see Figure 2).

## 2B – Check assumptions

We have already discussed the importance of modelling dependence in Section 2A. In addition, statistical models assume the response comes from a particular probability distribution with key properties including the response type (continuous, binary, counts) and it's mean-variance relationship. Examples include; the Gaussian (normal) distribution for continuous data like weight, which assumes constant variance; binomial distribution for presence/absence with a binary response and a quadratic mean-variance relationship; Poisson or negative binomial distribution for counts, like abundance, which assume the variance is equal to or greater than the mean respectively. Rather than fitting the data to the model, we should aim to develop a model that accounts for the characteristics of the data. If assumptions are not met, then the model does not account for the characteristics of the data, and researchers should endeavour to find a better fitting model. The violation of model assumptions will bias parameter estimates, for example by confounding location and dispersion effects (Warton et al., 2012); and increasing Type I errors (false positive findings; Queen et al., 2002), resulting in wrong conclusions from analyses. If a better model cannot be found (one may not exist or the correct model may not be implemented in code), then sensitivity analyses should be done. Sensitivity analysis is simply the process of trying alternative models which make different assumptions to assess if the conclusions of the analysis are robust to these changes (i.e. give consistent results).

Every model fitted must be checked and adjustments made before conclusions can be drawn. Assumptions are usually checked visually, by inspecting plots of residuals. A plot of residual vs. predicted values can show departure from linearity or the assumed mean-variance relationship. Plots of residuals over time or space would show the absence or presence of temporal or spatial autocorrelation, respectively. Normal quantile plots can diagnose departures from normality (if a normal distribution is used). The assumptions and sensitivity to violating the assumptions varies from model to model. For example, in a linear regression, the lack of normality is not as critical as constant variance, the violation of which may increase Type I error rates (Glass et al., 1972). As most distributions fit by ecologists are discrete (e.g. Poisson, negative-binomial, binomial), there are many choices of residuals. Quantile residuals, implemented in the `statmod` (Dunn & Smyth, 1996) and `DHARMa` (Hartig, 2020) packages are perhaps the most suited to checking assumptions for ecologists.

Sometimes we must make assumptions which cannot be checked to model our data. Two common scenarios are missing data, where it is necessary to make assumptions about the type of missingness (Nakagawa & Freckleton, 2008), and causal inference from observational data. Here, sensitivity analysis can help assess how sensitive results are to these assumptions.

# Principle 3: Emphasise effect sizes to replace statistical significance with ecological relevance

## 3A – Replace statistical significance with ecological relevance by emphasising effect sizes

Ecologists seek tools that allow them to say something about the ecological 'significance', or to use a better and less confusing term, assess the 'ecological relevance' of their findings (Martínez-Abraín, 2008). Since that is very hard, ecologists often settle for stating that their results are 'statistically significant'. While this is tempting, statistical significance is insufficient for the reporting of the results of ecological modelling because it does not provide an indication of effect sizes (Nakagawa & Cuthill, 2007).

Given a miniscule effect size in the population – say as an example, a 0.01% difference in the mean biomass between two water treatments – statistical significance can always be achieved by simply increasing the sample size sufficiently. This will not make a 0.01% difference any more ecologically relevant. The effect size (here, mean difference) is the quantity of interest, and must be prioritized in analysis and reporting, along with discussion on how ecologically relevant the measured effect size is, which must be based on expert knowledge.

Focussing on the estimation of effect sizes requires reporting and interpretation of confidence intervals (or equivalents when using other statistical approaches, such as credible intervals for Bayesian analysis). The width of these intervals indicates the uncertainty associated with an estimate. If you focus on estimation rather than significance testing, you can better capture the nuances of statistical analysis and interpretation.

In the swamp example, several effects were considered. The researchers were interested in what effect differing water treatments had on biomass, richness and composition, so analysis focused on differences between the control (high water level) and the other two levels (low and mid). For the fire treatment, researchers were interested in the interaction with the water treatment, that is, on differences of the effect of water level on biomass, species richness and abundance among burnt and unburnt mesocosms. For example, the authors hypothesised that biomass would decrease with decreasing water levels in unburnt treatments, and this effect would be even more pronounced in burnt ones.

## 3B – De-emphasize *p*-values

The *p*-value is defined as the probability, under the assumption of no effect or no difference in the population (null hypothesis), of obtaining a result equal to or more extreme than what was observed in the sample. This is a mouthful, and it is no wonder it is so commonly misinterpreted. Essentially this means *p*-values can indicate how incompatible the data are with a specified statistical model. A *p*-value is *not* the probability that the null hypothesis is true, nor is it an indicator of the size of the effect, or the probability that the data were produced by random chance alone, among other misinterpretations (Wasserstein & Lazar, 2016).

There have been calls for abandoning *p*-values and hypothesis testing altogether, both due to their ubiquitous incorrect interpretation, and due to previously mentioned problems like cherry-picking, and testing hypotheses which are known a-priori to be false. More commonly, statisticians recommend we stop dichotomizing *p*-values (into significant / non-significant; McShane et al., 2019) which not only confuses statistical and ecological significance, but is one of the causes of these bad practices like *p*-hacking. Statisticians also recommend de-emphasizing the importance of *p*-values. Current best practice is to report *p*-values as an addition to the more important quantities like effect size and confidence intervals, which also makes it harder to misinterpret the *p*-values. As Wasserstein et al., (2019, p. 2) recommend, when interpreting results it is best to "accept uncertainty. Be thoughtful, open, and modest".

When modelling changes in biomass over time in the swamp example, the researchers found 'evidence ($t_{341}$ = 3.093, P = 0.006) that differences in biomass between unburnt low and high water mesocosms increased over time, with biomass differences between high and low water mesocosms more than doubling (change = 2.183 ; 95% CI: 1.205 – 3.955) between Day 587 and Day 1261 of the experiment'. The confidence intervals suggest that the biomass difference is conceivably as small as a 20% increase (1.205) through to a quadrupling (3.955). The former might be viewed as being of modest ecological relevance while a doubling or more might be viewed as a large effect. The researchers additionally found 'no evidence of differences in biomass changes between high and medium water mesocosms ($t_{341}$ = 0.320, P = 0.945; change = 1.085; 95% CI: 0.597 – 1.971)'. The wide 95% confidence interval (0.597-1.971) suggests that the difference between high and medium water mesocosms could conceivably be halving (0.5) or doubling (2.0) of biomass. The large *p*-value and wide confidence interval may be a result of a too small sample size, or too large variance, or it could be that there is no difference between high and medium water mesocosms. The researchers have no way of knowing which of these factors was the cause of the large *p*-value, so they could not conclude that there is no difference between high and medium water levels.

# Principle 4: Report your methods and findings in sufficient detail so that your research is compelling and reproducible.

## 4A – Make it easy to reproduce your study findings

Replicability and reproducibility are important considerations when reporting your research. A result is reproducible when the same analysis steps performed on the same dataset consistently produces the same answer. A result is replicable when the same analysis performed on different datasets produces qualitatively similar answers(The Turing Way Community, 2023). As noted previously, ecology has very low rates of replication studies (Kelly, 2019), and most published studies do not provide sufficient detail about study design, data collection, and analysis for replication studies to even be attempted (Culina et al., 2020). Collaborative networks are perhaps a good way forward for increasing replication studies in ecology, like recent examples though the Nutrient Network, US Long-Term Ecological Research network, and Zostera Experimental Network. Apart from the scientific benefits of comprehensive reporting, the successful replication of your study will lend a lot more weight to your findings, and this should be your goal. You want to help anyone who attempts to replicate your work by providing all the information they might need.

If you have never included data and code with your publications, it can seem overwhelming, but it is not all or none, each step towards reproducibility is worthwhile.  Our advice follows that of a colleague (D. Falster, pers comm), who recommends you start small and work up to full reproducibility; first, writing the code so you yourself reproduce the results after some time has passed; second, by making sure a collaborator can reproduce them; and then finally it is only a small extra step to provide completely reproducible code in your publications. Some good practices, in approximate order of difficulty are:

- Always provide your complete data in a 'non-proprietary machine-readable format (i.e. not as hard scanned-in pages or PDF files);
- Keeping a research journal, noting any changes to study design or analysis, and why these were made, then reporting these in your methods and supplementary materials;
- Make sure your code is executable – i.e. your script file or Rmarkdown document should run from top to bottom without interruptions, exceptions and errors;
- Always supplement your papers with complete code to reproduce your results (and if using non-code based methods, provide a detailed workflow to trace your analytical steps);
- Where possible avoid using paid software and promote Open Source and freely available computational tools;
- Avoid using ephemeral and transient hosts to keep your code and data (e.g. personal websites, departmental web archives). Use free, publicly accessible and well-maintained repositories instead (e.g. Figshare, Dryad, Zenodo);
- When presenting code use version-control systems (such as Git) and rich documents integrating code with comments (e.g. by using Rmarkdown and R packages such as *knitr*);
- Declare packages used in the analysis and their versions (e.g. with the help of the `renv` package). If software/packages you use may change beyond being re-usable quickly consider packing your code and data into a self-contained package (for example, using Docker).

## 4B – Visualize model outputs to communicate results

As we have discussed in Principle 3, results should focus on effect sizes and confidence intervals to promote a focus on ecological relevance. One of the most effective ways to do this is to plot the *model outputs*. By model outputs, we mean the estimated effects and confidence intervals produced by the model. While some packages have inbuilt functions to do this, the `emmeans` package (Lenth, 2021) can plot model outputs from almost any commonly used R package. An alternative is to use the `predict()` function in most R packages to calculate predictions with uncertainty, then plot them manually. It is important to plot model outputs with uncertainty (i.e. confidence intervals), which is easily accomplished by using `CIs = TRUE` in `emmeans`, or `se.fit = TRUE` in many packages' `predict()` functions. The swamp study included plots created with `emmeans` (code below), then modified with `ggplot` (Wickham, 2016); model estimates and confidence intervals for richness over time are reproduced here (Figure 2).

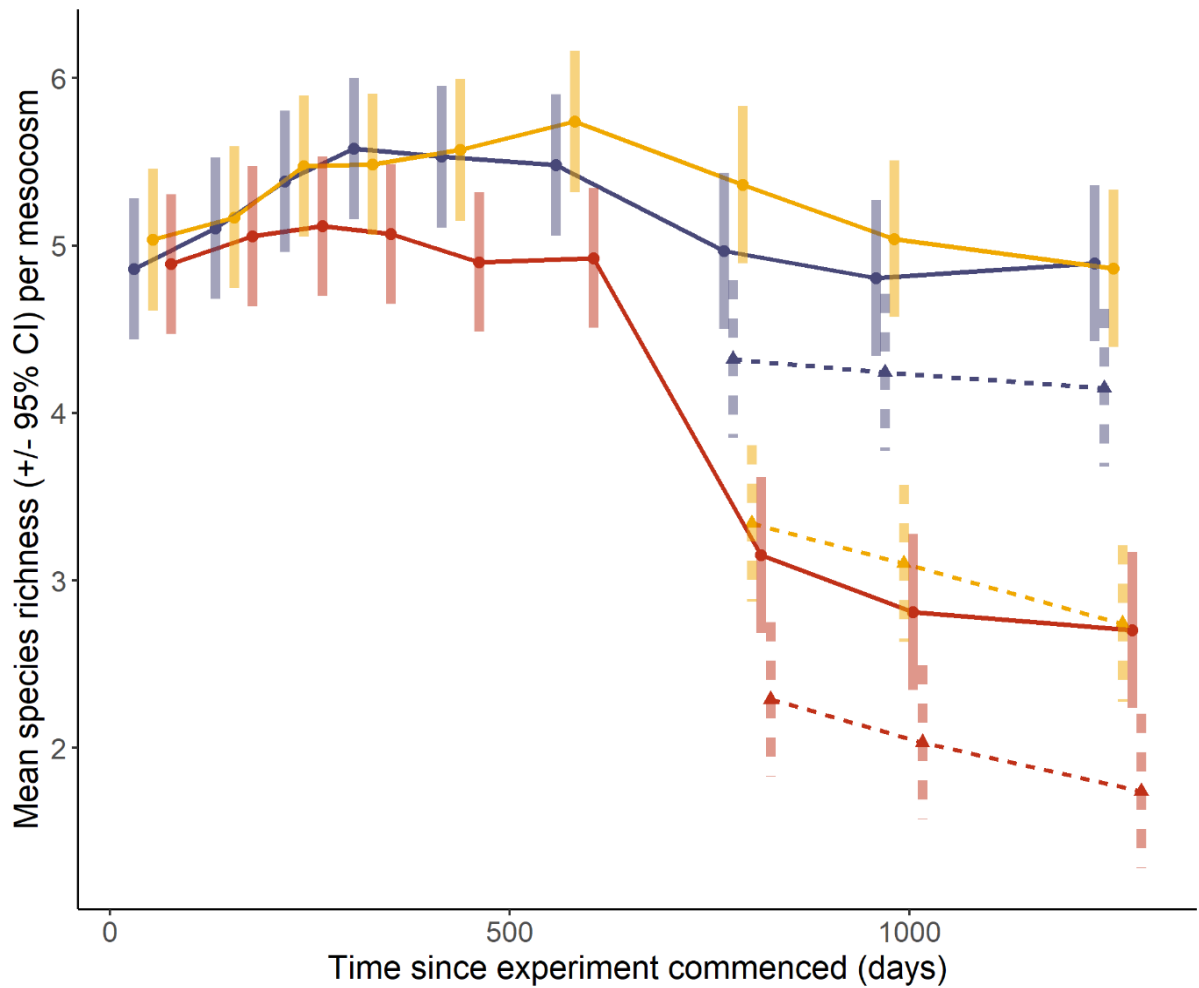

*Figure 2: Estimated marginal means (± 95% confidence intervals) for species richness over time (days since commencement) in treatment mesocosms. Treatment levels are high (━━) medium (━━) and low (━━) water availability and unburnt (━■━) and burnt (▪▪▪▲▪▪▪) fire treatment levels.*

## Discussion

This paper has proposed guidelines to enhance statistical methodology for ecological studies suitable for use by individual researchers and research teams. While outside of the scope of this work, undoubtedly there is more that can be done at a systemic level to encourage the adoption of stronger experimental design and reporting practices throughout the discipline, and the authors are aware of journals that are beginning to embed more robust methodological requirements into review processes. We acknowledge that it can be challenging for researchers to remain engaged with the complexities of statistical practice when deeply engaged in their specific areas of research, and it is our hope that clear guidance and critical engagement with statistical methods will help to build statistical competence and fluency to the benefit of the ecological research community.

## Author Contributions

## Funding

TJM acknowledges assistance by the NSW Government through its Environmental Trust (2018/SSC/0049) and Saving Our Species program. TAM thanks partial support by CEAUL (funded by FCT - Fundação para a Ciência e a Tecnologia, Portugal, through the project UIDB/00006/2020). RA is supported by National Research Foundation of South Africa (Grant No. 114696). ACW was supported by the São Paulo Research Foundation (Grant no. #17/16650–5). MMM was supported by the Christiane Nüsslein-Volhard Foundation. JAM was supported by the National Agency of Research and Innovation (ANII-Uruguay), and Computational Biodiversity Science and Services Program (Bios2-Canada). PP was supported by a UNSW Scientia Doctoral scholarship.

## Acknowledgements

The authors would like to thank everyone who participated in the vISEC2020 discussion group and subsequent discussions. We thank Glenda Wardle, Annemieke Drost, Nilanjan Chatterjee, Pedro Nicolau, Chloe Bracis, Teresa Neeman, Javier Seoane, Sarah Marley, Noa Rigoudy, Brenton Annan, Rebecca Groenewegen, Theresa O'Brien, Michelle Marraffini, Julie Vercelloni, Andrea Havron, Hayden Schilling, Amanda Hart, Christine Stawitz, Fabiana Ferracina, Andrew Edwards, Mick Wu, Gesa von Hirschheydt, Rick Camp, Alison Ketz, Julie Vercelloni, Sarah Saunders and Sarah Hasnain for their brainstorming contributions. Additionally, authors thank Frederic Gosselin for pointing out an important reference for testing implausible H0, and Daniel Falster for reproducibility advice.